\documentclass[journal=jacsat,manuscript=article]{achemso} 
\usepackage[latin9]{inputenc}
\usepackage{fontenc}
\usepackage{amssymb}
\usepackage{subfigure}
\usepackage{graphicx}
\usepackage{color}
\usepackage{units}
\usepackage{hyperref}
\usepackage{doi}
\usepackage{notes2bib}

\usepackage[version=3]{mhchem} 

\author{Samantha Micciulla}
\affiliation[Technische Universit\"{a}t Berlin]
{Stranski-Laboratorium, Institut f\"{u}r Chemie, Technische Universit\"{a}t Berlin, Strasse des 17. Juni 124, D-10623 Berlin, Germany}
\author{Pedro A. S\'{a}nchez}
\affiliation[Universit\"{a}t Stuttgart]
{Institut f\"{u}r Computerphysik, Universit\"{a}t Stuttgart, Allmandring 3, 70569 Stuttgart, Germany}
\altaffiliation{Current address: Faculty of Physics, Universit\"{a}t Wien, Boltzmanngasse 5, 1090 Wien, Austria}
\author{Jens Smiatek}
\affiliation[Universit\"{a}t Stuttgart]
{Institut f\"{u}r Computerphysik, Universit\"{a}t Stuttgart, Allmandring 3, 70569 Stuttgart, Germany}
\author{Baofu Qiao}
\affiliation[Universit\"{a}t Stuttgart]
{Institut f\"{u}r Computerphysik, Universit\"{a}t Stuttgart, Allmandring 3, 70569 Stuttgart, Germany}
\altaffiliation{Current address: Department of Material Science and Engineering, Northwestern University, Evanston, IL, 60208, USA}
\author{Marcello Sega}
\affiliation[Universit\"{a}t Stuttgart]
{Institut f\"{u}r Computerphysik, Universit\"{a}t Stuttgart, Allmandring 3, 70569 Stuttgart, Germany}
\altaffiliation{Current address: Institute for Computational Biological Chemistry, Universit\"{a}t Wien, W\"{a}hringerstr. 17, 1090 Wien, Austria}
\author{Andr\'e Laschewsky}
\affiliation[Fraunhofer-Institut f\"{u}r Angewandte Polymerforschung]
{Fraunhofer-Institut f\"{u}r Angewandte Polymerforschung (IAP), Geiselbergstr. 69, 14476, Potsdam-Golm, Germany}
\author{Christian Holm}
\email{holm@icp.uni-stuttgart.de}
\affiliation[Universit\"{a}t Stuttgart]
{Institut f\"{u}r Computerphysik, Universit\"{a}t Stuttgart, Allmandring 3, 70569 Stuttgart, Germany}
\author{Regine von Klitzing}
\email{klitzing@chem.tu-berlin.de}
\affiliation[Technische Universit\"{a}t Berlin]
{Stranski-Laboratorium, Institut f\"{u}r Chemie, Technische Universit\"{a}t Berlin, Strasse des 17. Juni 124, D-10623 Berlin, Germany}

\title[]{Layer-by-layer formation of oligoelectrolyte multilayers : a combined experimental and computational study}

\abbreviations{OEM, QCM-D, AFM, PDADMAC, PSS}
\keywords{Polyelectrolyte multilayers, short chains, atomistic simulation}

\begin{document}
\begin{abstract}
For the first time, the combination of experimental preparation and results of fully atomistic simulations of an oligoelectrolyte multilayer (OEM) made of poly(diallyl dimethyl ammonium chloride)/poly(styrene sulfonate sodium salt) (PDADMAC/PSS) is presented. The layer-by-layer growth was carried out by dipping silica substrates in oligoelectrolyte solutions and was modeled by means of atomistic molecular dynamics simulations with a protocol that mimics the experimental procedure up to the assembly of four layers. Measurements of OEM thickness, surface roughness and amount of adsorbed oligoelectrolyte chains obtained from both approaches are compared. A good agreement between simulated and experimental results was found, with some deviations due to intrinsic limitations of both methods. However, the combination of information extracted from simulations to support the analysis of experimental data can overcome such restrictions and improve the interpretation of experimental results. On the other hand, processes dominated by slower kinetics, like the destabilization of adsorbed layers upon equilibration with the surrounding environment, are out of reach for the simulation modeling approach, but they can be investigated by monitoring in situ the oligoelectrolyte adsorption during the assembly process. This demonstrates how the synergistic use of simulation and experiments improves the knowledge of OEM properties down to the molecular scale.
\end{abstract}

\section{Introduction}
The alternate adsorption of oppositely charged polyelectrolytes on charged surfaces---a thin film growth technique known as layer-by-layer (LbL) deposition\cite{1992-decher, Decher1997}---is a well established and straightforward route for the functionalization of solid substrates. The resulting systems are known as polyelectrolyte multilayers (PEMs) and they offer the possibility to create a surprising large variety of coatings by the proper choice of materials and assembly conditions. Since their introduction a huge number of systems have been created, comprising not only synthetic polymers but also biomaterials, micelles, inorganic components,\cite{Campbell2012,Findenig2013,Su2008} which stimulated numerous experimental and theoretical works devoted to their characterization.\cite{Schonhoff2003, Sukhishvili2005, Voigt2003} However, the fundamental study of PEMs faces difficulties that have limited a complete understanding of their properties and assembly mechanisms. For instance, experimental measurements at scales smaller than few tens of nanometers are challenging for many common techniques. Most analytical modeling approaches usually rely on strong, hard to prove approximations, and have serious difficulties to deal with the complex intermixing of the oppositely charged layers observed experimentally. In addition, most computer simulation studies have been based on top-down coarse-grained modeling approaches, characterized by rough approximations and a very limited connection with the microscopic properties of the system. For an overview about theoretical modeling and experimental work performed on PEMs and its open challenges the reader is addressed to recent literature\cite{Findenig2013, schlenoff2009, V.Klitzing2006, Volodkin2014, cerda12a} and references therein.

Besides the aforementioned lack of fundamental understanding, some potentially interesting approaches to the synthesis of PEMs with particular properties remain unexplored. For instance, although the preparation of PEMs makes use of polymers of different structural properties (linear, branched), charge density, functionality, molecular weights, very small chain lengths have not been used so far. The reason lies in the difficulty of trapping oligomeric complexes in the multilayer phase, due to the preferential formation of soluble complexes in bulk, which allows them to preserve high translational entropy.\cite{Sukhishvili2006,Sui2003}. However, the preparation of oligoelectrolyte multilayers (OEMs), is of fundamental importance for many reasons. From the perspective of applications, this strategy is expected to offer thin films with a different structure, porosity, elasticity and dynamics, but with the same surface chemistry as their equivalent long chain systems. From the point of view of theoretical modeling, and particularly concerning computer simulations, the use of oligomers significantly reduces the computational workload of atomistic models, making feasible the simulation of characteristics that remain unreachable to this approach when longer polyelectrolytes are involved.

Indeed, atomistic simulations may provide a proper understanding of the microscopic details underlying the assembly of OEMs, and may serve as a basis to bottom-up coarsening approaches properly grounded on the microscopic mechanisms of the system. In this context, we recently introduced an atomistic modeling approach for the formation of complexes of poly(styrene sulfonate sodium salt)/poly(diallyl dimethyl ammonium chloride) (PSS/PDADMAC) oligomers,\cite{Qiao2010} as well as the first extensive atomistic simulations for the adsorption on solid substrates of monolayers and bilayers of such oligoelectrolytes.\cite{qiao11a, qiao11b, qiao12a} These latter simulations represent the first steps in the growth of a PSS/PDADMAC OEM thin film, and have provided relevant details of the main microscopic mechanisms at the substrate-OEM interface region.

In this work we present the first experimental assembly and the first atomistic simulations of an OEM thin film obtained by the adsorption of up to four layers of PDADMAC/PSS oligomers on a negatively charged silica substrate. Our main goal here is to provide the first results on the mesoscopic properties of this novel system, showing the feasability of our experimental and atomistic simulation modeling approaches. In particular, the validation of our simulation model for the growth of PDADMAC/PSS OEMs by means of a comparison as much direct as possible with experimental results is a essential step for subsequent theoretical studies. However, this comparison is not straightforward. Although the low molecular weight of the oligoelectrolytes allows the atomistic simulation of a relatively large amount of molecules, it is still not possible to reach the lengths and time scales usually sampled in experimental measurements. Taking into account this limitation, we extended the system sizes explored in our previous simulations up to the boundaries of the maximum resolution power of the available experimental techniques. This has allowed us to extract from the atomistic simulation data a set of mesoscopic parameters that could also be determined experimentally with a reasonable accuracy. Therefore, we will focus our discussion on the comparison of the results provided by experiments and simulations for such selected mesoscopic properties, namely, the film thickness, the surface roughness and the amount of adsorbed oligoelectrolyte chains per layer.

The article is organized as follows: first, the common details of the system under study are presented; next, the experimental approach is described, including details about the materials, the experimental assembly procedure and the different techniques used for measuring the properties of the resulting films; then, a description of the computational modeling approach and the simulation protocol is provided; finally, both experimental and simulation results are presented and discussed, with a particular stress on the comparison between them.

\section{System under study}
\begin{figure}[h]
\centering
\includegraphics[width=.95\textwidth]{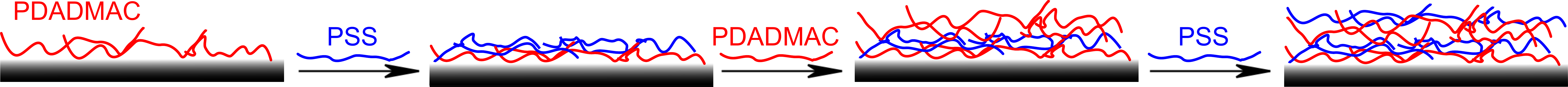}
\caption{Simplified sketch of the layer-by-layer (LbL) assembly of oligoelectrolyte multilayers from the alternate adsorption of PDADMAC/PSS chains.}
\label{fig:multilayer}
\end{figure}

In this work we study the formation and the mesoscopic structural properties of a four layers oligoelectrolyte multilayer which is composed of PDADMAC/PSS oligomers adsorbed on a silica substrate in presence of added salt ions. PDADMAC and PSS are polyelectrolytes frequently used for the assembly of multilayers. Since they are strong polyelectrolytes, their charge density---one unit charge per monomer for both macromolecules---is independent of the pH in solution. This reduces the parameters that control the conformations of the polymeric chains in solution and simplifies the modeling of the assembly mechanism. For this work, a degree of polymerization of about 30 monomers per chain was selected for both polyelectrolytes. The adsorption of the chains was chosen to be obtained from solutions at ionic strength $0.1M$ of monovalent salt (NaCl). Since silica substrates are known to have a negative surface charge at intermediate pH,\cite{Shin2002,Behrens2008} the multilayer assembly starts with the adsorption of PDADMAC chains. A simplified sketch of the structure of the four layers thin film is shown in Figure \ref{fig:multilayer}.

\section{Experimental approach}

\subsection{Materials}
Silicon wafers from Siltronic AG Siltron (Korea) were used as substrate for the adsorption of the oligoelectrolyte multilayers. Poly(sodium styrene sulfonate) PSS (6 520 Da, PDI=1.2) was purchased from Polymer Stardard Service (Mainz, Germany). Linear poly(diallyl dimethyl ammonium chloride) PDADMAC (5 000 Da, PDI=1.5) was synthesized at the Fraunhofer Institut f\"{u}r Angewandte Polymerforschung (IAP) (Potsdam-Golm, Germany). Sodium chloride NaCl (purity $>$99\%) was obtained from Merck (Germany). Polyelectrolyte solutions were prepared in Milli Q water (22 M$\Omega$cm$^{-1}$). All reagents were used as received without any further purification. 

\subsection{Preparation of polyelectrolyte multilayers}
Prior to oligoelectrolyte adsorption, silicon wafers were etched by a mixture of H$_2$SO$_4$:H$_2$O$_2$  1:1 v/v for 30 minutes \bibnote{Caution: piranha solution is highly corrosive, it should be handle with care and not be stored in closed containers}, rinsed thoroughly with Milli Q water and dried by N$_2$ stream. Oligoelectrolyte (OE) solutions were prepared by dissolving 0.01 mol(mono)/L in 0.1 M NaCl solution. Multilayers were prepared by dipping the substrate alternatively into oppositely charged OE solutions for 10 minutes. A rinsing in Milli Q water was performed between subsequent adsorption steps. No drying was done between depositions, therefore the samples with different number of layers were prepared on separated substrates. After preparation, the samples were gently blown dry with a stream of N$_2$ and stored in individual Petri dishes.

\subsection{Experimental methods}
The thickness of oligoelectrolyte multilayers was measured with a Multiscope Null-ellipsometer from Optrel GbR (Wettstetten, Germany). The instrument is equipped with a red laser (632.8 nm) and a PCSA (polarizer-compensator-sample-analyzer) setup. The angle of incidence was set to 60$^\circ$ for measurements in water. Light guides were used to drive the incident light at the substrate/water interface and avoid the reflection at liquid/air interface. The sample thickness was measured after multilayer preparation and drying with a N$_2$ stream. In order to determine the thickness in swollen state, each sample was left equilibrating in water for at least 15 minutes prior to measurement. The ellipsometric angles $\Delta$ and $\Psi$ were fitted with a four-layer box model, consisting of \emph{i)} water (refractive index $n = 1.33$), \emph{ii)} swollen polymer layer, \emph{iii)} silicon oxide SiO$_2$ ($n = 1.46$, thickness $d=1.5\ nm$) and \emph{iv)} silicon substrate Si ($n = 3.8858$). The refractive index for the swollen polymer layer was either obtained from measurements on thicker OEMs or calculated considering the ratio polymer/water within each deposited layer.

The surface roughness of OEMs was obtained from the height distribution of the surface normal to the substrate. The height profile was measured by Atomic Force Microscopy (AFM) using a Cypher Scanning Probe Microscope from Asylum Research (Germany). Surface scans were performed in Intermittent Contact Mode on samples swollen after preparation. Microcantilevers from Olympus TR400PSA for AC Mode in fluid were used, with dimensions (200 x 30 x 0.4) $\mu$m, spring constant 0.02 N/m and resonant frequency 10 kHz. A 4-sided silicon nitride tip is mounted on this cantilever, with radius of (20$\pm$5)nm. The scan rate was 0.5 Hz/s. Image analysis was done by using the software Igor Pro  (Version 6.3.4.1., Wave Metrics Inc.). The root-mean-square (rms) roughness was calculated from the standard deviation of the z-values obtained from a (1x1)$\mu$m$^2$ sample surface area. The reported values are the mean value of the roughness determined on four regions of (2x2)$\mu$m$^2$ scans.

The amount of adsorbed chains was calculated from the values of adsorbed mass. These values were in turn estimated from the frequency shift $\Delta$f measured by Quartz Crystal Microbalance with Dissipation (QCM-D) during the LbL assembly. A E1 QCM-D from Q-Sense (Sweden) was used, equipped with a silicon-coated quartz crystal (QSX303 from Q-Sense). The same OE solutions were used as for the dipping. Each solution was flowed in the preparation chamber for 10 min at a rate of 0.1 mL/min. Layer rinsing with Milli Q water was carried out for 5 min. From the measured frequency shift $\Delta$f, the total sensed mass $\Delta$m, including both oligomers and water, was estimated by the equation\cite{Sauerbrey1959}

\begin{equation}
 \Delta f=- \frac{2f_0^2}{A\sqrt{\rho_q\mu_q}}\Delta m,
\label{eq:sauerbrey}
\end{equation}
where $f_0$ is the fundamental frequency of quartz (4.95 MHz for AT cut quartz sensors), $A$ the piezoelectrically active area, $\rho_q$ the quartz density (2.648 g cm-3) and $\mu_q$ the quartz shear modulus (2.947$\cdot$10$^{-11}$ gcm$^{-1}$ s$^{-2}$ ). The contribution of water to the sensed mass was subtracted and the surface coverage  was calculated as number of adsorbed chains per cm$^2$.

\section{Computational approach}
\label{sec:sim}
\begin{figure}[h]
\centering
\includegraphics[width=.35\textwidth]{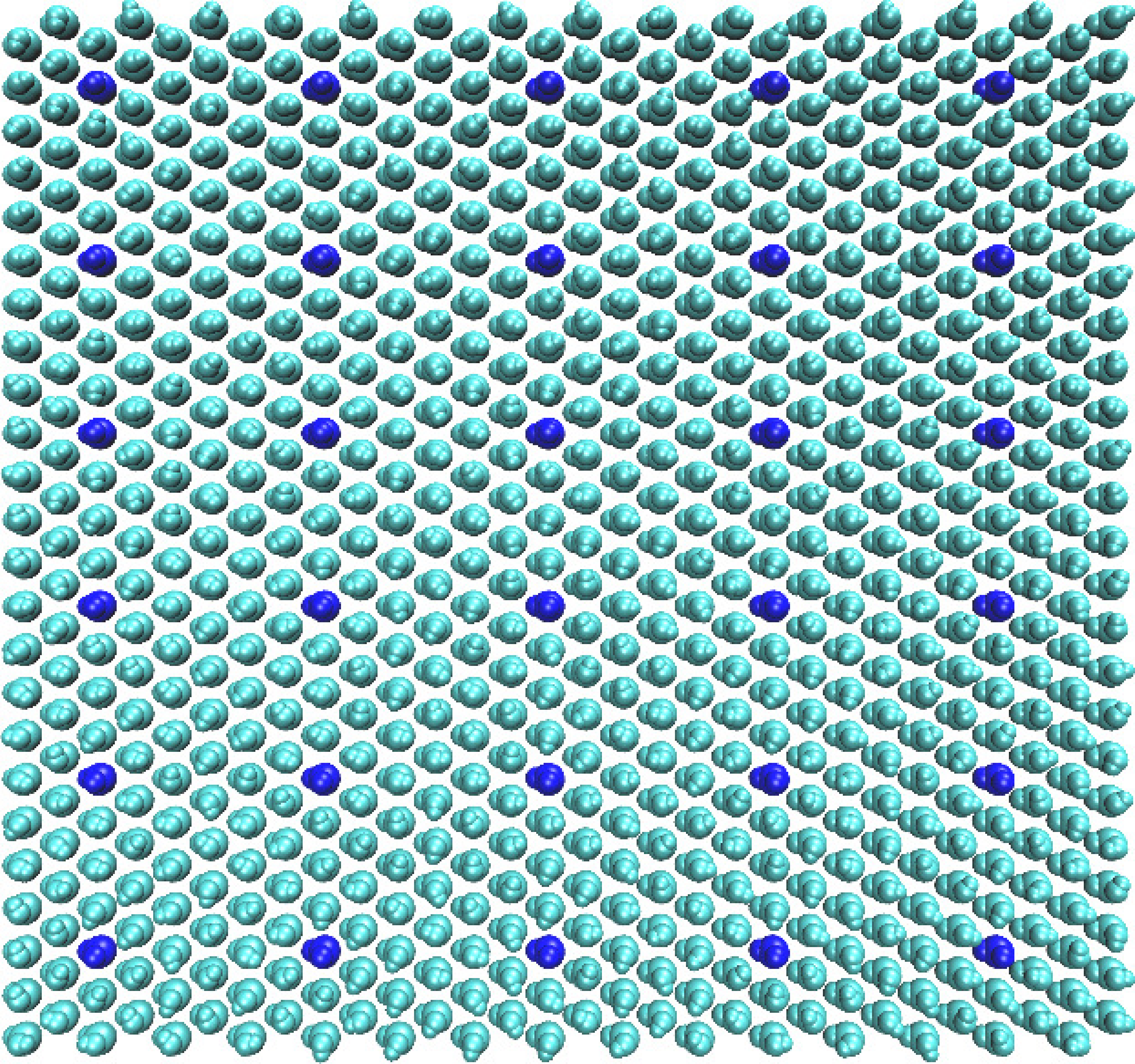}
\caption{Overhead view of the distribution of hydroxyl groups (light spots) and surface charges (dark spots) in the interfacial layer of the silica substrate used for the atomistic simulations.}
\label{fig:simsetup}
\end{figure}

We simulated the growth of (PDADMAC/PSS)$_2$ OEMs by means of fully atomistic classical molecular dynamics (MD). The simulations were performed with the package Gromacs 4.5.3.\cite{hess08a} The interactions of the oligomers were modeled with the OPLS-AA force field.\cite{jorgensen96a} Details of the parametrization of this force field for our system can be found in Ref. \cite{Qiao2010}. The SPC/E water model\cite{berendsen87a} was chosen to simulate the solvent due to its capability to reproduce accurately the hydration behavior and the dielectric constant of water.\cite{hess06b, hess06c} Geometry constraints were applied to water molecules according to the SETTLE algorithm.\cite{miyamoto92a} The silica substrate was modeled atomistically as a four layers sheet silicate structure with dimensions in the $X$ and $Y$ directions of $13.1650\ nm$ and $12.1608\ nm$, respectively. Such dimensions were selected to fit the crystalline structure of the substrate to lateral periodic boundaries, which were used in all the simulations, while keeping an aspect ratio as low as possible. In addition, the chosen substrate dimensions are close to the nominal range of the radius of the AFM scan tip apex, as indicated by its manufacturer. The substrate surface was made hydrophilic by replacing the uppermost oxygen atom of every silicate tetrahedron on the interfacial layer with a polar hydroxyl group (\ce{\bond{-}OH}). Such hydroxyl groups were allowed to freely rotate around an axis normal to the surface,\cite{ho11a} in order to avoid the simulation artifacts found when their orientations are fixed.\cite{qiao11b} Finally, a distribution of surface charges was obtained by skipping one in every four of the hydroxyl group allocations, just leaving exposed the corresponding \ce{(SiO_2)^-} group instead. This procedure led to a surface number density of $4.3\ nm^{-2}$ for the hydroxyl groups and a surface charge of $\sigma=-3\cdot 10^{-20}\ C/nm^2$, which is in good agreement with the surface charge measured on silica in comparable experimental conditions.\cite{Shin2002,Behrens2008} Figure \ref{fig:simsetup} shows the distribution of hydroxyl groups and surface charges over the substrate surface. Our simple approach for the allocation of the surface charges, which corresponds to a discrete distribution in a regular square lattice, is based on the assumption that the local microscopic details of the surface charge have a negligible impact on the average mesoscopic properties of our system. This hypothesis is supported by the results of a recent computational study on the mesoscopic dynamics of electro-osmotic flows with charged walls\cite{2009-smiatek} and is in agreement with previous computational works.\cite{qiao11a, qiao11b, qiao12a}

The simulation protocol employed here is basically the same that we developed in our previous works on the atomistic simulation of oligoelectrolyte monolayers and bilayers.\cite{qiao11a, qiao11b, qiao12a} This protocol tries to mimic the LbL experimental deposition process at time scales accesible to atomistic simulations. Its main steps are the following: the adsorption of one given layer begins by placing a equilibrated solution of the corresponding oligoelectrolyte chains---the dipping solution---close to the adsorbing surface. Such dipping solution consists of 20 chains of oligoelectrolytes with a length of 30 monomers, and their 600 respective counterions (\ce{Cl^-} for PDADMAC and \ce{Na^+} for PSS), 99286 water molecules and a concentration of $0.1M$ of added salt. The latter was obtained by adding 192 \ce{Na^+} and \ce{Cl^-} ions, with a final equilibrium volume given by the substrate lateral dimensions, $13.1650  \times 12.1608\ nm^2$, and a height of roughly $19.9\ nm$. The dipping solution was initially prepared with the package Packmol 1.1.1\cite{martinez09a} and equilibrated for $50\ ns$ at a pressure of $1\ bar$ in the $Z$ direction, applied with a semi-isotropic Parrinello-Rahman barostat.\cite{1981-parrinello, 1983-nose} The whole system---the substrate and previous layers plus the dipping solution---was then equilibrated in four steps: first, a relaxation of the initial configuration was allowed by energy minimization with the steepest descent algorithm. In the three subsequent steps, MD simulations in the NVT ensemble were performed using the Noose-Hoover thermostat at a reference temperature of $T=298K$ and a characteristic time $\tau=0.5\ ps$. A neighbor searching cut-off of $1.3\ nm$ was used. Van der Waals interactions were simulated with Lennard-Jones potentials shifted to smoothly decay to zero at distances larger than $0.9\ nm$. Electrostatic interactions were calculated by means of the particle mesh Ewald method (PME),\cite{darden93a, essmann95a} with a direct space cut-off of $1.3\ nm$ and a Fourier grid spacing of $0.125\ nm$. In order to obtain a pseudo-2D Ewald summation, required by the slab geometry of the system, the Yeh-Berkowitz correction to the force and potential\cite{1999-yeh} was applied, and a region of twice the height of the initial system was kept empty in the upper part of the simulation box to avoid the artifacts associated to this method. The first NVT equilibration step ran for $100\ ps$ with a time step of $1\ fs$. In the next equilibration step, which ran as well for $100\ ps$, the covalent bonds of the hydrogen atoms were constrained with the LINCS algorithm,\cite{hess97a} and the time step was increased to $2\ fs$. Finally, in the last equilibration step all the covalent bonds were constrained with the LINCS method, and the simulation ran for $300\ ns$ with a time step of $2.5\ fs$. During this latter step, the number of oligoelectrolyte chains adsorbed to form the new layer was monitored and found to reach a constant value, thus indicating that a effective equilibrium was achieved. Once the adsorption of the new layer was completed, a rinsing of the film was performed by simply removing the supernatant solution with the remaining unadsorbed chains and ions. The rinsed film resulting from this procedure was just formed by the substrate and the adsorbed oligoelectrolyte layers, the surrounding water molecules up to the maximum height of the adsorbed oligoelectrolytes, and the counterions required to keep an overall charge neutrality. Such rinsed system was then used as the new adsorbing surface for the next dipping solution. The complete sequence of dipping-adsorption-rinsing procedures was carried out for four times by alternating PDADMAC and PSS disolutions for the dipping step, thus producing a simulated four layers OEM. In order to obtain a minimal statistical sampling of the structure of the four layers system, three completely independent runs were conducted.

The measurements of the properties of the simulated films were carried out in separate equilibration and production runs intended to mimic the experimental conditions of pure water swollen films. With this purpose, after the adsorption and rinsing of every new layer, the empty space left by the removal of the supernatant dipping solution was refilled with a box of pure water, previously equilibrated in a separate simulation. The same simulation steps used for the adsorption of every layer were applied again to equilibrate this swollen system, except for the very last step, in which the running time was limited to just $100\ ns$. This choice is based on the assumption that the main changes during the equilibration of the swollen system correspond to the redistribution of waters and counterions, which have a rather faster dynamics than the oligoelectrolyte chains during their adsorption process. Therefore, we have not tried to simulate any slow kinetic effect---like eventual instability and desorption of OE-OE complexes from the film surface that may occur during the experimental rinsing process---due to its still prohibitive computational cost for fully atomistic simulations. Finally, the atomistic trajectories were sampled at a rate of $0.1\ ns$ during the latter $100\ ns$ step. However, only the measurements obtained in the last $50\ ns$ were taken into account for the final results. 

As a final remark, due to the relatively high number of atoms involved in these simulations---above $4\cdot 10^5$---the total amount of computing power spent has been of approximately $2.2\cdot 10^6$ cpu-hours.

\section{Results and discussion}
In this section the measurements of the film thickness, surface roughness and amount of adsorbed chains obtained from experiments and simulations are presented and compared.

 
\subsection{Film thickness}
The thickness of the simulated films was calculated from the swollen equilibrated samples by computing the average height of the film surface with respect to the substrate adsorption plane. In particular, the film surface was defined by discretizing the plane parallel to the substrate as a rectangular lattice of $50\times50$ cells. For each given cell, the maximum height of any enclosed atom belonging to either PDADMAC or PSS chains was taken as the film surface height at the position of the cell. According to this spatial discretization and to the lateral dimensions of the film, the calculation of the film surface had a maximum resolution of approximately $0.26\ nm$. For the calculation of the thickness, in addition, the regions with a zero height---\textit{i.e.}, the regions in which the substrate was not covered by any polymer atom, either in close contact with it or as part of overhanging structures---were not used for the averaging. Taking into account that in the simulated system water molecules tend to fill completely the polymer free regions, this calculation strategy was intended to match the experimental ellipsometry measurements of the film thickness, in which the fine structural details are averaged out in the measured optical parameters.
Specifically, the experimental determination of the film thickness, $d$, was obtained by measuring the ellipsometric angles, $\Delta$ and $\Psi$, using fixed values for the refractive index of the film, $n$. In general, ellipsometry measurements allow the determination of both refractive index and thickness of a sample from the change e of phase and amplitude of the reflected light. Unfortunately, the reflection from ultrathin films---like the ones studied here---only the change of phase can be determined. Therefore, either the refractive index or the thickness has to be estimated in some way in order to determine the other parameter. For thickness measurements, the approach usually used to solve this problem consists of making analogous measurements on thicker films, determining $n$ and $d$ from them, and then using the so obtained refractive index as a fixed single value to calculate the thickness of thinner films. We applied this strategy by carrying out independent measurements on thicker, (PDADMAC/PSS)$_{20}$ films, obtained with the same preparation protocol described above, and consequently assuming that the resulting estimation of $n$ applies equally to our whole (PDADMAC/PSS)$_{2}$ films. Nevertheless, in this case we can alternatively take advantage of the simulation data in order to obtain an estimation of the ratio water/polymers present in every layer of the film, from which an estimation of the refractive index per layer can be obtained. By applying this refined estimation of the effective refractive index, some improvement in the accuracy of the measured thickness was reasonably expected.

The single refractive index for (PDADMAC/PSS)$_2$ OEMs estimated from the measurements performed on thicker multilayer was $n=1.458$. For the calculation of a refractive index per layer from the simulation data, the water and aggregated oligoelectrolyte mass distribution profiles along the direction of growth of the multilayer were computed for the simulated films in order to estimate the variation of the water mass fraction as a function of the layer number. The result of this calculation shown a significant drop of the relative water content in the region corresponding to the second layer, where the presence of PSS chains is higher, with respect to the regions corresponding to the first (PDADMAC), third (PDADMAC) and fourth (PSS) layers. In particular, this drop was of approximately a 30\% with respect to the first layer, and 25\% with respect to the third and fourth layers. Qualitatively, the drop of water content in the second layer with respect to the first and third layers is supported by the experimental observations of a lower hydration in PSS layers in front of the PDADMAC ones. \cite{IturriRamos2010, Feldoto2010} Unfortunately the statistical error of such results, as estimated from the three available simulation runs, was found to be rather large, specially for the third and fourth layers, in which the fluctuations were above the 50\%. This evidences that the sampling statistics achieved in our simulations for this paramater were relatively poor. Therefore, this calculation of the water mass fraction per layer from the simulation data should be considered just a rough estimation to support a qualitative estimation of layer thickness from the experimental data. Consequently, the values of water contents per layer used for the experimental data were selected as follows: for the first layer, a reference value of water content of 40\% was taken from data reported in literature,\cite{Dodoo2011} which is similar to our own swelling measurements carried out on high molecular weight (PSS/PDADMAC) PEMs (data not reported here). From this reference value, the variations determined from the simulation data were applied to the next layers. Therefore, for the second layer the assumption of the aforementioned decrease of 30\% led a drop of the water content to around 10\%. For the third and fourth layers, the aforementioned increase of water content of 25\% with respect to the second layer was assumed as well, bringing the relative amount of water to 35\%. The values of the refractive index obtained with such estimations were, from first to third layer, $n=1.462$, $1.528$ and $1.473$, respectively.

\begin{figure}[h]
\centering
\includegraphics[width=.5\textwidth]{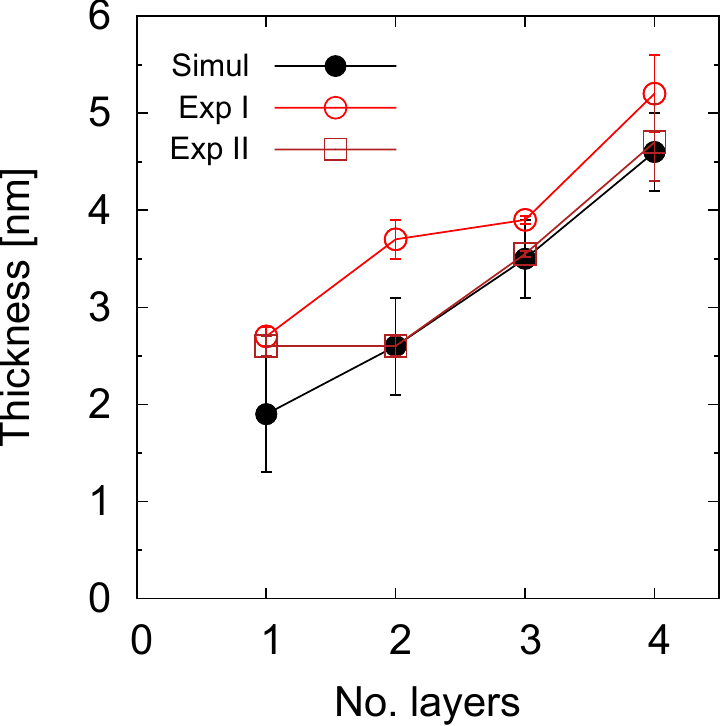}
\caption{Thickness as a function of the number of layers obtained from simulations (filled circles) and experimental ellipsometry measurements on samples swollen in water (open circles and squares). Even data correspond to PDADMAC termination, odd data to PSS termination. Experimental values correspond to two different estimations of the refractive index: a fixed value $n=1.458$ for the whole film (open circles, Exp I), and a set of layer dependent values, from first to fourth $n=\lbrace 1.458,\ 1.462,\ 1.528,\ 1.473 \rbrace$, corresponding to the water mass fraction estimated from simulations (squares, Exp II).}
\label{fig:thickness}
\end{figure}

The growth of the film thickness with the number of deposited layers, determined directly from simulations and from both experimental ellipsometry approaches, is shown in Figure \ref{fig:thickness}. Considering the low values involved in this measurement, a reasonably good agreement with the simulation data is observed for both experimental results. However, for the case of the experimental thickness obtained with a single refractive index for the whole film (Exp I in Figure \ref{fig:thickness}), a systematic shift to higher values can be noticed. This could be due to an overestimation of the water content within some regions of the film, which leads to the observed deviation towards apparently thicker films. To prove that, we tested the sensitivity of our ellipsometry measurements to the accurate estimation of the refractive index by calculating the range of thickness values obtained when $n$ varies within a relatively narrow interval. Table 1 in the supporting material shows the results: it is remarkable how a significant variation in the measured thickness, from $8.6$ to $3.2\ nm$, is observed when $n$ is changed just from $1.40$ to $1.55$.

However, Figure \ref{fig:thickness} also demonstrates that there is an excellent agreement between the direct simulation results and the ellipsometry measurements when the relative content of water per layer obtained from the simulations is used to calculate $n$ per each layer (Exp II in Figure \ref{fig:thickness}). This agreement is even more remarkable if we consider the poor statistics on which such estimation was based. This result illustrates how computer simulations may support experimental measurements, particularly when they are carried out close to the limits of their maximum resolution power.

\subsection{Surface roughness}
\begin{figure}[h]
\centering
\includegraphics[width=.5\textwidth]{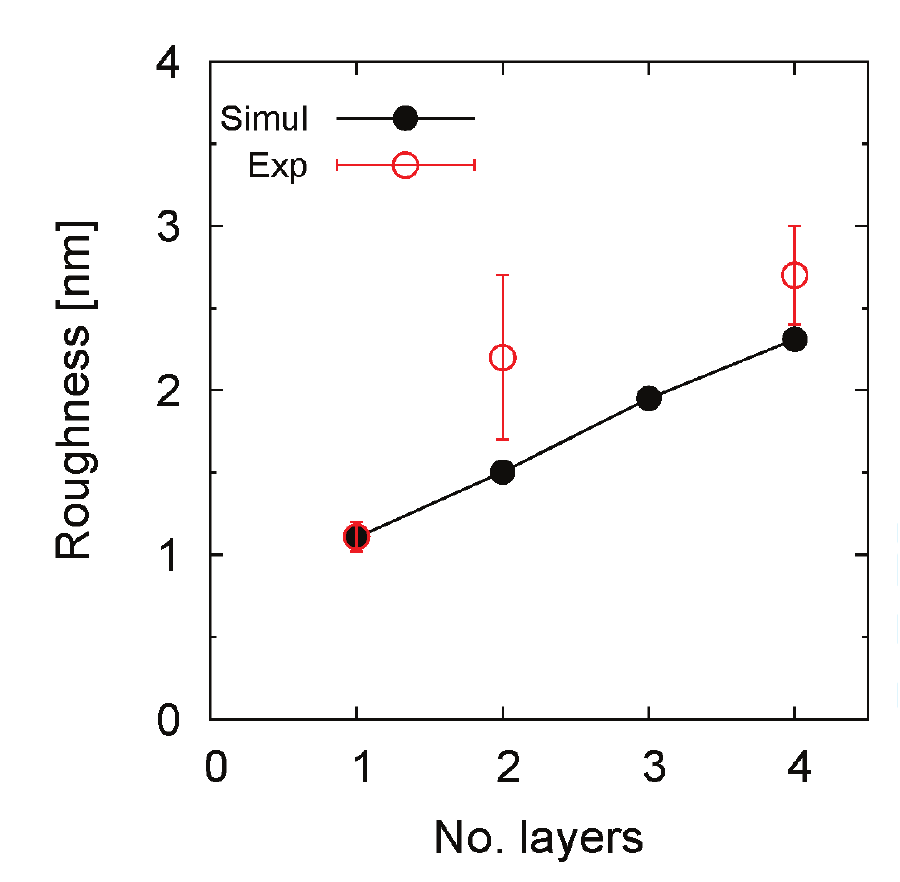}\\
\caption{Root mean square roughness of PDADMAC/PSS OEMs measured on simulated (filled circles) and experimental samples (open circles), as a function of the number of adsorbed layers. Experimental data were obtained from AFM measurements carried out on multilayers swollen in water. Even data correspond to PDADMAC termination, odd data to PSS termination.}
\label{fig:2}
\end{figure}

The rms roughness of (PDADMAC/PSS) OEMs calculated from simulated and experimental OEM surface profiles is reported in Figure \ref{fig:2} against the layer number. The simulated data were extracted from the same surface profiles used for the determination of the film thickness. Due to the relatively small size of the surface area sampled in the simulations, all the data obtained from the three independent runs were aggregated to calculate the rms deviation, hence no error estimation is reported in this case. As mentioned in the \textit{Experimental Approach} section, the experimental data were determined from AFM measurements and averaged on four different (1x1)$\mu$m$^2$ regions on each sample.

As shown in Figure \ref{fig:2}, the data obtained from the two approaches present a reasonable qualitative and quantitative agreement: in both cases, the surface roughness grows with the number of deposited layers. However, the experimental measurements show moderately higher averages than the corresponding to the simulations. Such disparities are not surprising for AFM scans carried out on films with this low number of deposited layers. It is known that the spatial resolution of AFM instruments is limited by the size of the tip apex, so the profile of the surface at scales smaller than such size is poorly sampled. In addition, the simulation data shows a strong variation of the substrate coverage within the explored range of deposited layers, growing continously from roughly a 30\% for the first layer to a 80\% for the fourth one. If we assume a comparable substrate coverage variation for the experimental films, a significant and presumably inhomogeneous impact on the accuracy of the corresponding AFM measurements should be expected. Even though we have some evidence that the presence of a regular charge pattern in our model surface should not have a critical impact on the results presented here, its influence on the conformation of the adsorbed chains, especially of the first layer, cannot be completely ruled out, and will be the subject of further investigations.

\subsection{Adsorbed oligoelectrolyte chains}
The adsorption of a polyelectrolyte on a solid substrate is a complex process and it can be schematically described by three main steps: I) initial adsorption, characterized by a strong mass uptake, driven by electrostatic attraction of oppositely charged species and entropic gain by counterions release; II) chain rearrangement, when the adsorbed chains rearrange in a more stable conformation, increase the number of complexation sites and entanglements; III) rinsing, when a polymer-free solution, either pure water or salt solution, is flowed on the layer to remove the excess of PE chains and counterions prior to the next adsorption cycle. Among the change of mass measured during each of the mentioned steps, the adsorbed mass after rinsing is the most important one, since it determines the film growth over a full set of adsorption-rinsing cycles. This is exactly the quantity that was intended to be reproduced by the computer simulations. Instead, no representative data for the actual adsorption and rinsing kinetics is expected to be obtained from such approach due to the strong limitations in the simulated time scales and the simplifications introduced in the simulation protocol. On the other hand, experimental measurements of the polymer adsorbed mass can be made for every one of the three adsorption steps described above by means of the QCM-D technique. In particular, QCM-D provided measurements of the frequency shift $\Delta f$, that were used together with Equation \ref{eq:sauerbrey} to obtain an estimation of the variations of total mass in the OEM. From such an estimation, the number of oligoelectrolyte chains per unit area was calculated for each layer after substracting the contribution of swelling water to the total sensed mass. The relative water content per layer estimated from the simulation data for the thickness measurements---from first to fourth: 40\%, 10\%, 35\% and 35\%, respectively---is assumed here again. 

\begin{figure}[h]
\centering
\includegraphics[width=.48\textwidth]{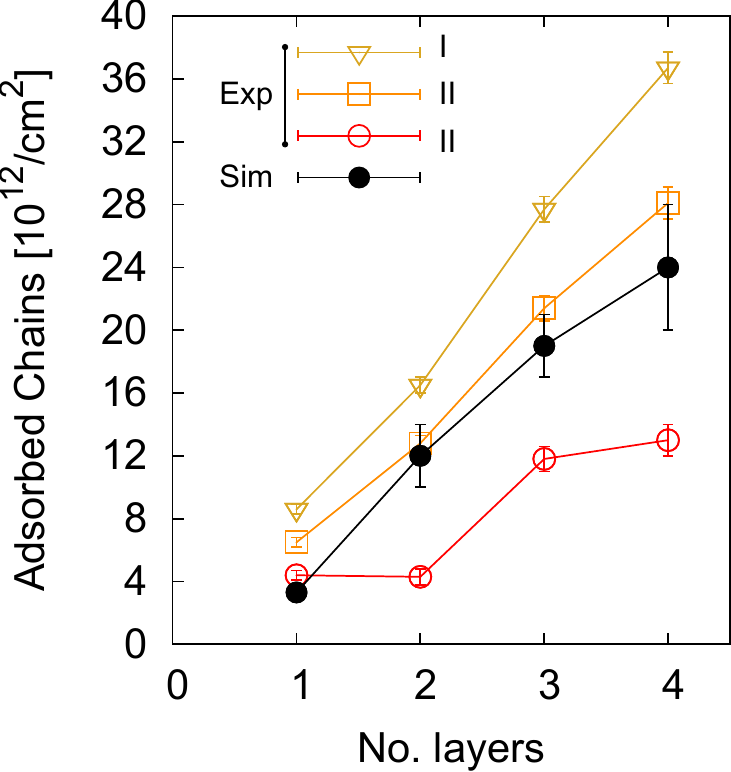}
\caption{Accumulated number of OE chains per unit area as a function of the layer number from simulated data (filled circles) and calculated from the experimental adsorbed mass (open circles). The siulated data correspond to the adsorbed mass after water rinsing, \textit{i.e.}, at the end of a full adsorption-rinsing cycle. The experimental data were calculated from the sensed mass measured afted the initial adsorption (triangles), chain rearrangement (squares) and rinsing step (circles).}
\label{fig:chains}
\end{figure}

Figure \ref{fig:chains} shows the accumulated number of adsorbed chains per layer for both simulated and experimental data. If the data relative to the number of adsorbed chains after rinsing are analyzed (Exp III), it can be noticed that there is a good agreement between simulated and experimental data for the first PDADMAC layer, as well as for the increment after the adsorption of the third layer. Nevertheless, for the case of PSS layers, second and fourth, the variations of the experimental values are significantly lower than the ones corresponding to simulations.

The reported trends illustrates some important aspects of the LbL assembly of PDADMAC/PSS OEMs:
\begin{itemize}
\item[i)] the adsorbed mass decreases systematically from the initial adsorption step (I) to the rinsing step (III) for all layers. This means that, after a strong uptake driven by electrostatic interactions and entropic gain upon counterions release, the equilibration of adsorbed chains in the layer leads to some mass loss. The rinsing by pure water brings an additional mass loss, due to the removal of the excess of oligoelectrolyte chains and counterions.
\item[ii)] The water rinsing after PSS deposition causes a strong mass removal, leading to very small layer increment. This is partially due to layer dehydration during rinsing, as proved by atomistic simulations and supported by the decrease of energy dissipation measured by QCM-D (Figure S1 in the Supporting Material), but it also proves a pronounced layer destabilization
during rinsing promoted by the abrupt change of osmotic pressure from OE solution to salt-free water. The possibility of a significant layer degradation within the experimental time scales is eased by the absence of a strongly entangled structure, which is unlikely due to the short length of OE chains and to the higher mobility of oligoelectrolyte complexes compared
to the long chain ones.
\end{itemize}

Unfortunately, the aforementioned limitations prevent the observation of these effects---pres\-umably of slow kinetics nature---in our atomistic simulations, and no net decrease in the amount of adsorbed chains was handled to be reproduced. Nevertheless, these experimental observations are crucial to understand the effect of slower kinetic processes on the effective growth of OEM layers.

\subsection{Conclusions}
In the present work, poly(diallyl dimethyl ammonium chloride)/poly(styrene sulfonate sodium salt) oligoelectrolyte multilayers were characterized by atomistic simulations and experiments. The comparison between thickness, roughness and number of adsorbed chains obtained by the two approaches shows a good agreement within our range of confidence. It was also demonstrated that the experimental limitation arising from a poor accessibility to structural details on a molecular scale can be overcome by taking advantage of atomistic simulations. In this case, when the refractive index of the swollen layer was calculated by considering the swelling ratio extracted by the atomistic simulation, the agreement between the results was improved. Nevertheless, atomistic simulations are still unable to reproduce molecular processes dominated by slower kinetics, like slow chain rearrangements and destabilizations, but those processes have a crucial role in determining the layer growth. Therefore the use of QCM-D in this work to monitor in situ the change of mass after the initial strong uptake added important information to the investigation of the system dynamics.

The use of the water content inside the layer to assist the determination of ellipsometric layer thickness and the investigation of kinetic processes on experimental time scales are the examples that demonstrate the benefit of the common effort of simulations and experiments to characterize the properties of complex macromolecular systems on different molecular levels.

\begin{acknowledgement}
We acknowledge the \textit{Deutsche Forschungsgemeinschaft (DFG)} within the Priority Program SSP 1369 for financial support. We also thank the following projects and organizations for the computational resources: bwGRiD \bibnote{BwGRiD (http://www.bw-grid.de), member of the German D-Grid initiative, funded by the Ministry for Education and Research (Bundesministerium fuer Bildung und Forschung) and the Ministry for Science, Research and Arts Baden-Wuerttemberg (Ministerium fuer Wissenschaft, Forschung und Kunst Baden- Wuerttemberg).}, and Paderborn Center for Parallel Computing, Universit\"at Paderborn.
\end{acknowledgement}


\providecommand*\mcitethebibliography{\thebibliography}
\csname @ifundefined\endcsname{endmcitethebibliography}
  {\let\endmcitethebibliography\endthebibliography}{}

\pagebreak
\section{Supporting material}
\renewcommand\thefigure{S\arabic{figure}}    
\setcounter{figure}{0}

\begin{table}
\centering
\begin{tabular}{c | c}
\multicolumn{2}{c}{(PDADMAC/PSS)$_2$}\\
\hline
\emph{n} (fixed) & \emph{d} [nm] (fitted)\\
1.40 & 8.6 \\
1.41 & 7.6 \\
1.42 & 6.8 \\
1.45 & 5.3 \\
1.46 & 4.9 \\
1.47 & 4.6 \\
1.50 & 3.9 \\
1.55 & 3.2
\end{tabular}
\caption{Values of thickness obtained from the fit of ellipsometric angles $\Delta$ and $\Psi$ of (PDADMAC/PSS)$_2$ OEMs for different fixed values of refractive index \emph{n}.}
\label{table:indexes}
\end{table}

\begin{figure}[h]
\centering
\includegraphics[width=.5\textwidth]{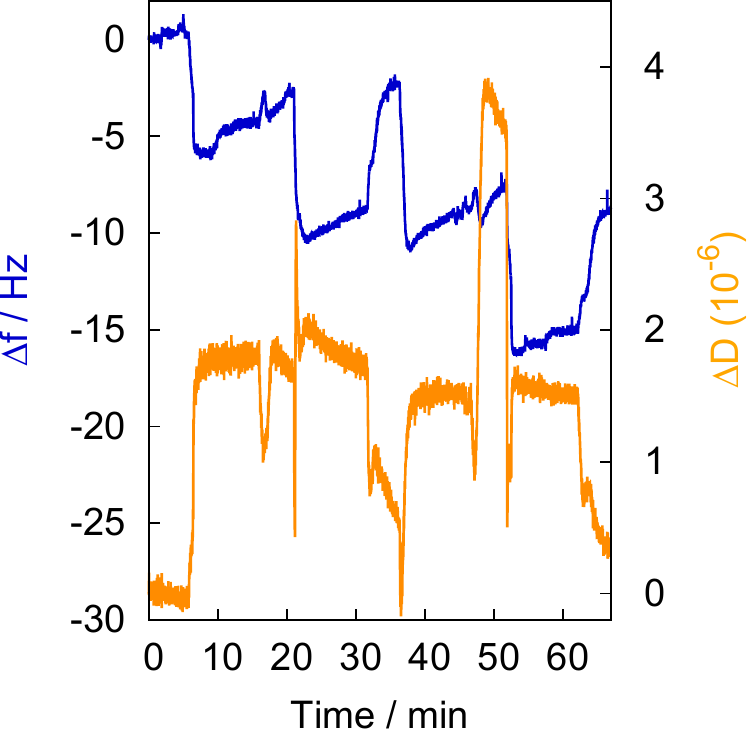}
\caption{Frequency shift ($\Delta$f) and dissipation energy ($\Delta$D) recorded by QCM-D during the layer-by-layer growth of PDADMAC/PSS oligoelectrolyte multilayers on silicon-coated quartz crystal. Flow rate of OE solutions 0.01 mol(mono)/L in NaCl 0.1 M was 0.1 mL/min. Salt-free water was used for rinsing between subsequent oligomer depositions.}
\label{fig:S1}
\end{figure}

\begin{figure}[h]
\centering
\includegraphics[width=.9\textwidth]{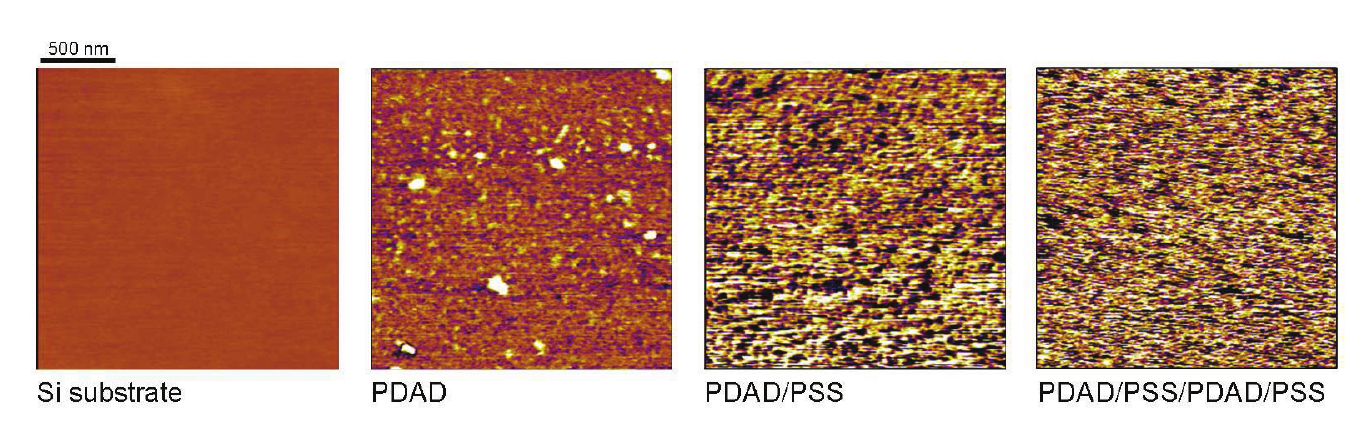}
\caption{Atomic force microscopy (AFM) scans performed on samples swollen in water. The scan dimensions are (2x2)$\mu$m$^2$}
\label{fig:S2}
\end{figure}
\clearpage


\begin{mcitethebibliography}{42}
\bibitem[Not()]{Note-1}
Caution: piranha solution is highly corrosive, it should be handle with care
  and not be stored in closed containers\relax
\mciteBstWouldAddEndPuncttrue
\mciteSetBstMidEndSepPunct{\mcitedefaultmidpunct}
{\mcitedefaultendpunct}{\mcitedefaultseppunct}\relax
\EndOfBibitem
\end{mcitethebibliography}
\end{document}